\documentclass{article}
\pdfoutput=1
\usepackage{graphicx}

\newcommand\blfootnote[1]{%
  \begingroup
  \renewcommand\thefootnote{}\footnote{#1}%
  \addtocounter{footnote}{-1}%
  \endgroup
}

\newcommand{\aptaweb}{AptaBlocks Web Interface}

\title{AptaBlocks Online - A web-based toolkit for the in-silico assembly of RNA complexes}
\author{Jan Hoinka\,$^{\dagger}$ \and Yijie Wang\,$^{\dagger}$ \and Teresa M. Przytycka\,$^{*}$\\\\
National Center for Biotechnology Information, NLM, NIH, \\Bethesda, Maryland, USA 20894\\ \texttt{przytyck@ncbi.nlm.nih.gov} }
\date{\today}

\begin{document}

\maketitle
\blfootnote{$^\ast$To whom correspondence should be addressed.}
\blfootnote{$^\dagger$ Contributed equally}

\abstract{\textbf{Summary:} The \aptaweb\ is focused on providing graphical, intuitive, and platform independent access to AptaBlocks, an experimentally validated algorithmic approach for the in-silico design of oligonucleotide sticky bridges. The availability of AptaBlocks online to the nucleic acid research community at large makes this software a highly effective tool for accelerating the design and development of novel oligonucleotide based drugs and other biotechnologies. \\\\
\textbf{Availability:} The \aptaweb\ is freely available at \\ \texttt{www.ncbi.nlm.nih.gov/CBBresearch/Przytycka/index.cgi\#aptablocks} \\
}

\maketitle
\section{Introduction}
Oligonucleotides are rapidly emerging as competitive and highly efficient alternatives to traditional therapeutics and are increasingly employed in a growing array of biotechnological applications \cite{rossi2017, ellington2010, zu2014, giangrande2010}. These include, but not limited to next-generation biosensors \cite{williams2012, dobson2014, soh2013} and, more recently, in \textit{in vitro} and \textit{in vivo} delivery systems for individualized medicine approaches which frequently face the challenge of delivering a drug across a biological membrane to reach their target. \\

A large number of these technologies rely on a multi-stage architecture in which several biomolecules are combined together to form a biologically functional unit. Personalized drug therapy particularly benefits from such a modular model as these approaches typically consist of a case-specific and exchangeable delivery molecule joint with a target-specific therapeutic.\\ %and this multi-stage approach allows for the the design of the therapeutical component and of the delivery component independently from each other.

While designing a molecule to cross a particular membrane has proven challenging, aptamers, small RNA or DNA molecules explicitly designed to bind a target of interest with high specificity and affinity, have established themselves as a prominent choice for this task. 
As a case in point, one of the earlier approaches successfully created a cell internalizing delivery system \textit{in vitro} by coupling an anti-prostate specific membrane antigen aptamer %, that had previously been shown to bind to prostate tumor cells, 
with an anti-Lamin A/C siRNA for the inhibition of gene expression \cite{levy2006}. More recently, \textit{in vivo}, prostate cancer cells expressing prostate-specific membrane antigen (PSMA) have been successfully targeted with optimized aptamer-siRNA chimeras resulting in pronounced regression of PSMA-expressing tumors in athymic mice after systemic administration \cite{giangrande2009}. In addition, an aptamer-cargo conjugate is currently being assessed as a targeted delivery system for pancreatic cancer in vivo \cite{rossi2016}. \\

The success of the above mentioned approaches hinges on the ability to effectively conjugate the delivery agent with the cargo. While previous generations of oligo-cargo complexes have relied on conjugation techniques such as streptavidin based bridges and similar methods \cite{levy2006}, more recent approaches have opted for a more flexible nucleotide hybridization technique. This so called sticky bridge approach consists of two short and complementary sequence strands which are covalently bound to the delivery and cargo oligonucleotide respectively and are expected to hybridize, hence forming the desired complex in solution.\\ % The advantage of this approach is that it enables the design of flexible 'mix and match' strategies where a single cargo can be delivered to an arbitrary number of targets using just one sticky bridge. Conversely, a single delivery agent can be reused to carry multiple therapeutics to the same target. 

The specific sequence of such bridges however must be carefully designed such that its stands do not interact with neither the delivery component nor the therapeutic as to form undesired dimers. This process has traditionally been a laborious and costly trail-and-error process involving the synthesis of an initial sticky bridge based on personal expertise, followed by experimental testing for hybridization, subsequent fine-tuning of the sequence, and  iteration of these steps until the desired binding strength and stability of the complex is achieved.\\

To improve this process, and to accelerate the design of RNA-based drug delivery systems, we recently developed  AptaBlocks \cite{wang2018}, a computational method to design optimized sticky bridges while minimizing the probability of undesired dimer formation (see Supplementary Section 2). While the utility of AptaBlocks for novel, mutli-component oligotherapies is clearly evident, our program is currently available as a command line interface (CLI) tool only, possibly restricting its accessibility to the broader biotechnology community.\\

Here, we present the \aptaweb, a fully featured, platform independent version of AptaBlocks that can be accessed from any web browser. In contrast to its CLI counterpart, the \aptaweb\ guides the user through an intuitive, graphical step-by-step process allowing for the generation of sticky bridges that are optimized for up to five cargoes simultaneously.\\

\begin{figure}[!tpb]%figure1
\centerline{\includegraphics[width=\linewidth]{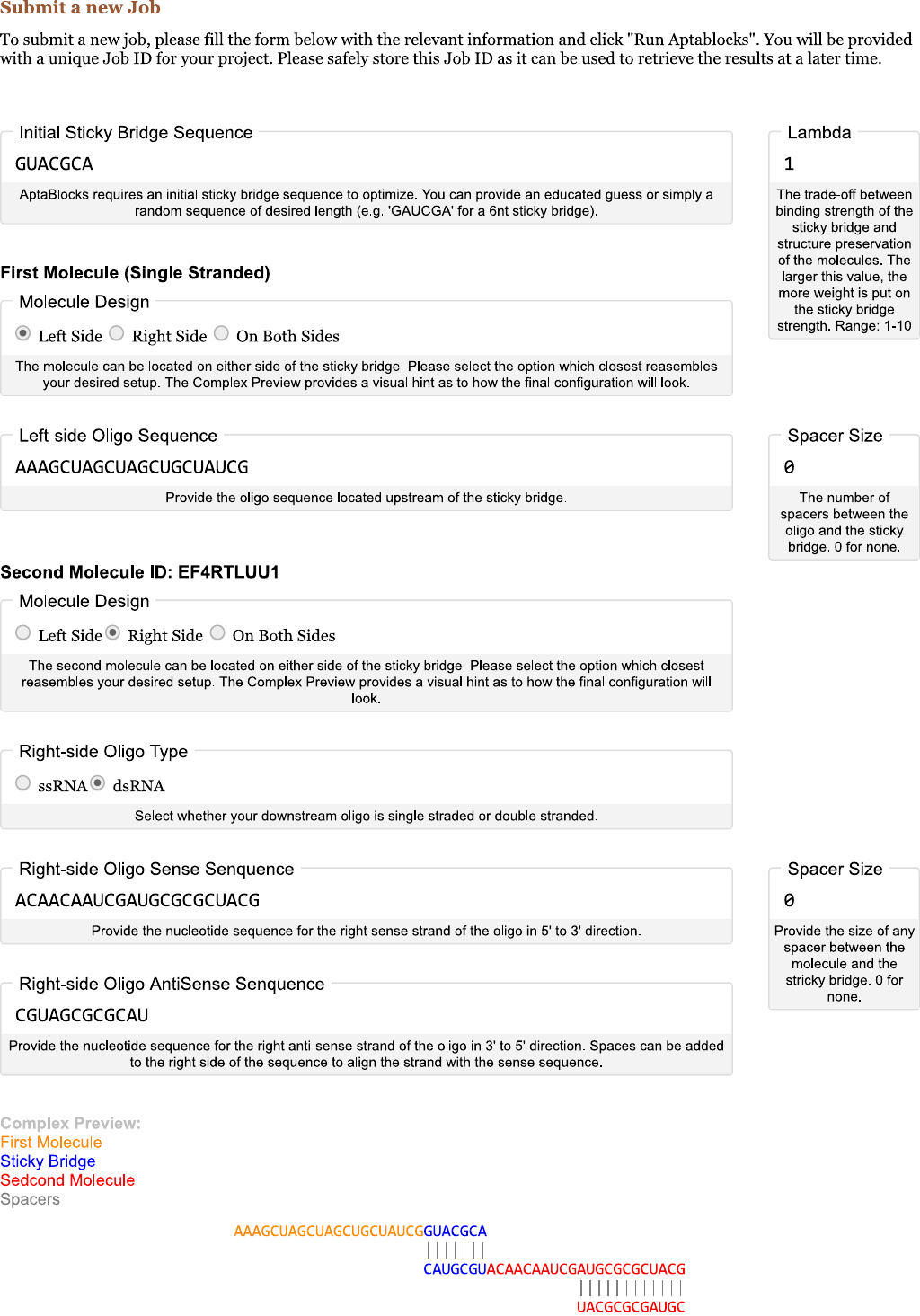}}
\caption{Screen capture of the \aptaweb. Shown are exemplary values for the different parameters a user is guided through in order to design an optimal sticky bridge for the desired input oligos. Each option is accompanied with a detailed description regarding the requirement and purpose of this property. In addition, the user is presented with a preview of the complex in real time.}\label{fig:01}
\end{figure}

\section{Web Server}
The web server is partitioned into two parts dedicated to parameter input and results presentation. In what follows, we provide a brief overview of the structure and functionality of these segments.\\

\textbf{Parameter Input: } The \aptaweb\ consists of two main sections for the exploration of sticky bridges. The first section allows the user to retrieve any previous jobs from the server by entering the unique id number assigned to every new job and provided to the user upon submission.
Depending on the validity of the job id, the user will either be informed that no such job exists, or is presented with the results screen as detailed in the Results Section. Alternatively, should the corresponding job still be running, the user is presented with a waiting screen displaying the job id, the status of the job (i.e. queued, running, completed, or failed), and the elapsed time on the server since submission in seconds. \\

The second section allows for the design of a new sticky bridge with one delivery oligonucleotide (First Molecule) and up to five cargoes (Second Molecules) as shown in Figure \ref{fig:01}. Each cargo is given a unique identifier for later differentiation. \\

The \aptaweb\ requires an initial sticky bridge sequence to optimize which can consist of an educated guess or a random sequence of desired length (e.g. 'GAUCGA' for a 6nt sticky bridge). A non-default value in the range of 1 to 10 for Lambda, the parameter controlling the trade-off between the binding strength of the sticky bridge and structure preservation of the molecules, can be entered. The larger this value, the more weight is put on the sticky bridge strength during the optimization process.\\ 

Next, the properties of the first molecule can be defined. First, the location of the sticky bridge on the oligo strand is selected which can either be positioned on the five prime end, the three primer end, or somewhere in the center of the sequence. Depending on this choice the user is prompted to provide the oligo sequence located upstream, downstream, or on both sides of the sticky bridge. In addition, the number of desired 3C spacers, between the oligo and the sticky bridge can be defined for each of the selected sequences. These elements consist of one or more three-carbon molecules that are used to incorporate a short spacer arm into an oligonucleotide and computationally treated as incapable of pairing with any nucleotide.\\

The options for the second molecule(s) are analogous to that of the first one with the additional ability to select the cargo sequence to be either single stranded (ssRNA) or double stranded (dsRNA). Furthermore, up to five molecules can be configured to be optimized simultaneously through the "Add Oligo" button. \\

Finally, for each of the delivery-cargo complexes, a live preview of the complex is created in order to provide the user with a visual aid of the molecule to be optimized. \\

\textbf{The Results Section:} \label{sec:results} Upon completion of the optimization process with AptaBlocks, or upon retrieval of a previously completed job, the optimized sticky bridge sequence is presented to the user along with a summary of the parameter chosen on the initial submission form. For each of the second molecules, the probability of preserving the structure of the delivery oligo, the probability of preserving the structure of the cargo, the probability of forming undesired dimers, as well as the free energy of the sticky bridge is presented to the user. A visual preview of the corresponding delivery-cargo complex is also provided.\\

\section{Conclusion}
A multitude of current and future nucleic acid based, personalized drug therapies as well as other biomedical approaches such as next-generation biosensors require the formation of molecular complexes from two or more oligonucleotides in order to perform their intended action. Finding the specific nucleotide sequence to effectively link the individual components together via sticky bridges has traditionally remained a tedious, wet-lab based task that is very much dependent on personal expertise. This process has recently been complemented with a systematic, computational approach that significantly streamlines and accelerates the design of the sticky bridge and adds the ability optimize the latter to be compatible with a multitude of molecules. This tool, known as AptaBlocks, however has until now only been available in command line form, possibly limiting its availability to the extended nucleic acid research community. The \aptaweb\ closes this gap by providing AptaBlocks online though a distributed, high-availability computation grid with an appropriate graphical interface that is not only platform independent, but also intuitive and highly user friendly to use.

\section*{Acknowledgements}
We would like to thank the NCBI C++ Toolkit team for developing and maintaining the NCBI GRID framework and particularly Vakatov D, Sadyrov R, and Fukanchik S for their excellent support during the development stages of the web interface.  \vspace*{-12pt}

\bibliographystyle{unsrt}
%\bibliography{references}

\end{document}